\documentclass[aps,prl,10pt,showpacs,floatfix,twocolumn,amsmath,amssymb,superscriptaddress,portrait]{revtex4-1}
\usepackage{pdfpages}
\usepackage{pgffor}

\usepackage{graphicx}
\usepackage{bm,soul}

\usepackage{hyperref,color}
\usepackage[all]{hypcap}

\newcommand{\Ang}{\AA$^{-1}$}

\newcommand{\banisbo}{Ba$_3$NiSb$_2$O$_9$}
\newcommand{\basrnisbo}{Ba$_{2.5}$Sr$_{0.5}$NiSb$_2$O$_9$}

\newcommand{\XQE}{\ensuremath{\chi''(Q,E)}}
\newcommand{\SQE}{\ensuremath{S(Q,E)}}

\newcommand{\figdir}{figmain/}

\begin{document}

\title{Evidence for a spinon Fermi surface in the triangular \texorpdfstring{$S=1$}{S=1} quantum spin liquid \texorpdfstring{Ba$_3$NiSb$_2$O$_9$}{Ba3NiSb2O9}}

\author{B.~F\aa k}
\email{fak@ill.fr}
\affiliation{Institut Laue-Langevin, CS 20156, 38042 Grenoble Cedex 9, France}

\author{S.~Bieri}
\email{samuel.bieri@alumni.epfl.ch}
\affiliation{Institute for Theoretical Physics, ETH Z\"urich, 8099 Z\"urich, Switzerland}
\affiliation{Laboratoire de Physique Th\'eorique de la Mati\`ere Condens\'ee, CNRS UMR 7600, Universit\'e Pierre et Marie Curie, Sorbonne Universit\'es, 75252 Paris, France}

\author{E.~Can\'evet}
\affiliation{Institut Laue-Langevin, CS 20156, 38042 Grenoble Cedex 9, France}
\affiliation{Laboratory for Neutron Scattering and Imaging, Paul Scherrer Institut, 5232 Villigen, Switzerland}
\affiliation{Laboratoire de Physique des Solides, Universit\'e Paris Sud 11, CNRS UMR 8502, 91405 Orsay, France}

\author{L.~Messio}
\affiliation{Laboratoire de Physique Th\'eorique de la Mati\`ere Condens\'ee, CNRS UMR 7600, Universit\'e Pierre et Marie Curie, Sorbonne Universit\'es, 75252 Paris, France}

\author{C.~Payen}

\author{M.~Viaud}

\author{C.~Guillot-Deudon}
\affiliation{Institut des Mat\'eriaux Jean Rouxel, CNRS UMR 6502, Universit\'e de Nantes, 44322 Nantes Cedex 3, France}

\author{C.~Darie}
\affiliation{Institut N\'eel, CNRS, Univ.\ Grenoble Alpes, Bo\^ite Postale 166, 38042 Grenoble Cedex, France}

\author{J.~Ollivier}
\affiliation{Institut Laue-Langevin, CS 20156, 38042 Grenoble Cedex 9, France}

\author{P.~Mendels}
\affiliation{Laboratoire de Physique des Solides, CNRS, Universit\'e Paris-Sud, Universit\'e Paris-Saclay, 91405 Orsay Cedex, France}

\date{December 22, 2016}

\begin{abstract}
Inelastic neutron scattering is used to study the low-energy magnetic excitations in the \mbox{spin-1} triangular lattice of the $6H$-B phase of \banisbo. We study two powder samples: \banisbo\ synthesized under high pressure and \basrnisbo\ in which chemical pressure stabilizes the $6H$-B structure. The measured excitation spectra show broad gapless and nondispersive continua at characteristic wave vectors. Our data rules out most theoretical scenarios that have previously been proposed for this phase, and we find that it is well described by an exotic quantum spin liquid with three flavors of unpaired fermionic spinons, forming a large spinon Fermi surface.
\end{abstract}


\maketitle

Quantum spin liquids (QSLs) are exotic phases of condensed matter where the ground state evades ordering as a consequence of strong quantum fluctuations, frustration, or topological effects. QSLs are related to resonating-valence-bond states \cite{AndersonHsu87_PRL.58.2790}, and they exhibit fascinating properties such as long-range entanglement and fractional excitations \cite{Balents10_Nature_464_199, *Lee08_Science_321_5894, MoessnerRamirez06, MendelsBert_CRP_16}. The natures of such ground states are hotly debated questions, both in candidate materials \cite{Han2012, Fak12_PRL_109_037208, Shimuzi03_PRL_91_107001, 131dmit3, FuImaiLee_Science05} and in theoretical models \cite{Lauchli11_PRB.83.212401, Depenbrock12_PRL.109.067201, Iqbal13_PRB.87.060405}, in particular concerning the existence of an excitation gap. Theoretically, a plethora of distinct and interesting possibilities for QSL phases has been classified \cite{Wen02_PRB.65.165113, Messio13_PRB_87_125127, Bieri16_PRB.93.094437}. To date, spin liquids have mainly been sought for in low-dimensional spin $S=1/2$ systems, where quantum fluctuations are strongest. A pressing topic is therefore the existence of QSLs and their nature in systems with higher values of spin \cite{Liu10_PRB.82.144422}.

The $6H$-B phase of \banisbo\ \cite{Balicas11Ni_PRL.107.197204} is of particular interest in this context. The Ni$^{2+}$ ions form a frustrated triangular lattice of $S=1$ spins. No sign of magnetic ordering is observed in the magnetic susceptibility down to $T=2$~K \cite{Balicas11Ni_PRL.107.197204}, in the specific heat down to $0.35$~K \cite{Balicas11Ni_PRL.107.197204}, or in muon spin rotation ($\mu$SR) measurements down to $T=0.02$~K \cite{Quilliam16_PRB.93.214432}, while the Curie-Weiss constant of $\theta_{\rm CW} = -76$~K indicates dominant antiferromagnetic interactions \cite{Balicas11Ni_PRL.107.197204}. Strikingly, when $T \ll | \theta_{\rm CW}|$, the compound shows a large linear term in the specific heat, $\gamma = 168$~mJ/mol~K$^2$, and a finite magnetic susceptibility \cite{Balicas11Ni_PRL.107.197204}. Such a metallic behavior in this strong Mott insulator suggests the presence of gapless coherent quasiparticles, possibly due to the emergence of a Fermi sea of fractional spinons. Evidence of gapless spin excitations are also found in recent NMR and $\mu$SR measurements~\cite{Quilliam16_PRB.93.214432}.

Several scenarios have been discussed so far to explain the intriguing properties of \banisbo. The $S=1$ spin of the Ni$^{2+}$ ions can be fractionalized into three \cite{Liu10_PRB.81.224417} or four \cite{Xu12_PRL.108.087204} fermionic spinons, resulting in rather different, but plausible QSL states: A chiral $\mathbb{Z}_2$ QSL with spinon Fermi surface \cite{Serbyn11_PRB.84.180403, *Serbyn13_PRB.88.024419, Bieri12_PRB_86_224409} or a time-reversal symmetric $\mathbb{Z}_4$ QSL with quadratic spinon bands touching \cite{Xu12_PRL.108.087204} have been proposed. Nematic three-dimensional spin liquids resulting from a bosonic fractionalization of spin have also been put forth \cite{HwangYBK13_PRB.87.235103}. Other proposals include the proximity to a quantum critical point as a consequence of fine-tuned inter- and intralayer exchanges, without the formation of a spin-liquid ground state \cite{Chen12_PRL.109.016402}.

In this paper, we study powder samples of the $6H$-B structure of \banisbo\ using inelastic neutron scattering~(INS) in order to bring clarity to these theoretical proposals. Broad gapless and nondispersive spin excitation continua are observed at three characteristic wave vectors. Strikingly, our wave-vector resolved data rule out most of the previous proposals for the magnetic low-temperature phase. We find that the INS data is well described by a U(1) quantum spin liquid with three flavors of spinons, forming a large spinon Fermi surface. This exotic spin $S=1$ QSL state preserves full spin-rotation and time-reversal symmetry, as well as all symmetries of the triangular lattice.

The $6H$-B phase of \banisbo\ reported in Ref.\ \cite{Balicas11Ni_PRL.107.197204} crystallizes in the $P6_3mc$ space group with  two Ni$^{2+}$ ions at the $2b$ Wyckoff site, which form triangular layers of $S=1$ spins with quenched orbital moments, stacked such that a Ni$^{2+}$ ion in one layer sits above the center of the triangle formed by Ni$^{2+}$ ions in the layer below. These layers are separated by nonmagnetic Sb layers, and appear well decoupled. We  synthesized under pressure a 0.7~g powder sample of this $6H$-B phase of \banisbo, as described in Ref.~\cite{Darie2016166}. However, such a small quantity is hardly sufficient for detailed INS studies. We therefore made a larger 6.1~g powder sample of \basrnisbo, where chemical pressure via partial Ba/Sr substitution stabilizes the $6H$-B structure \cite{suppMat}. Rietveld analyses of x-ray diffraction data from this \basrnisbo\ sample collected at room temperature were performed using the published $P6_3mc$ or $P6_3/mmc$ crystal structures of $6H$-B \banisbo\ as a starting model \cite{Darie2016166}. As in pure $6H$-B \banisbo, the best refinement was obtained for the $P6_3/mmc$ model \cite{suppMat}. The related structural questions, discussed in Ref.\ \cite{Darie2016166}, concern essentially the stacking of the triangular layers, and are of little importance for the two-dimensional magnetic properties dealt with in the present work.

Our magnetic susceptibillity measurements of \basrnisbo\ show an absence of magnetic order down to $T=2$~K and a Curie-Weiss temperature of $\theta_{\rm CW}=-80$~K \cite{suppMat}, in close agreement with earlier measurements on $6H$-B \banisbo\ \cite{Balicas11Ni_PRL.107.197204}. This suggests that the partial replacement of Ba with Sr does not change the magnetic properties of the compound. Assuming nearest-neighbor (NN) Heisenberg interactions, the Curie-Weiss temperature implies an antiferromagnetic NN exchange of $J_1 \sim 20$~K with the convention of counting each bond once.

\begin{figure}
\includegraphics[width=0.49\columnwidth]{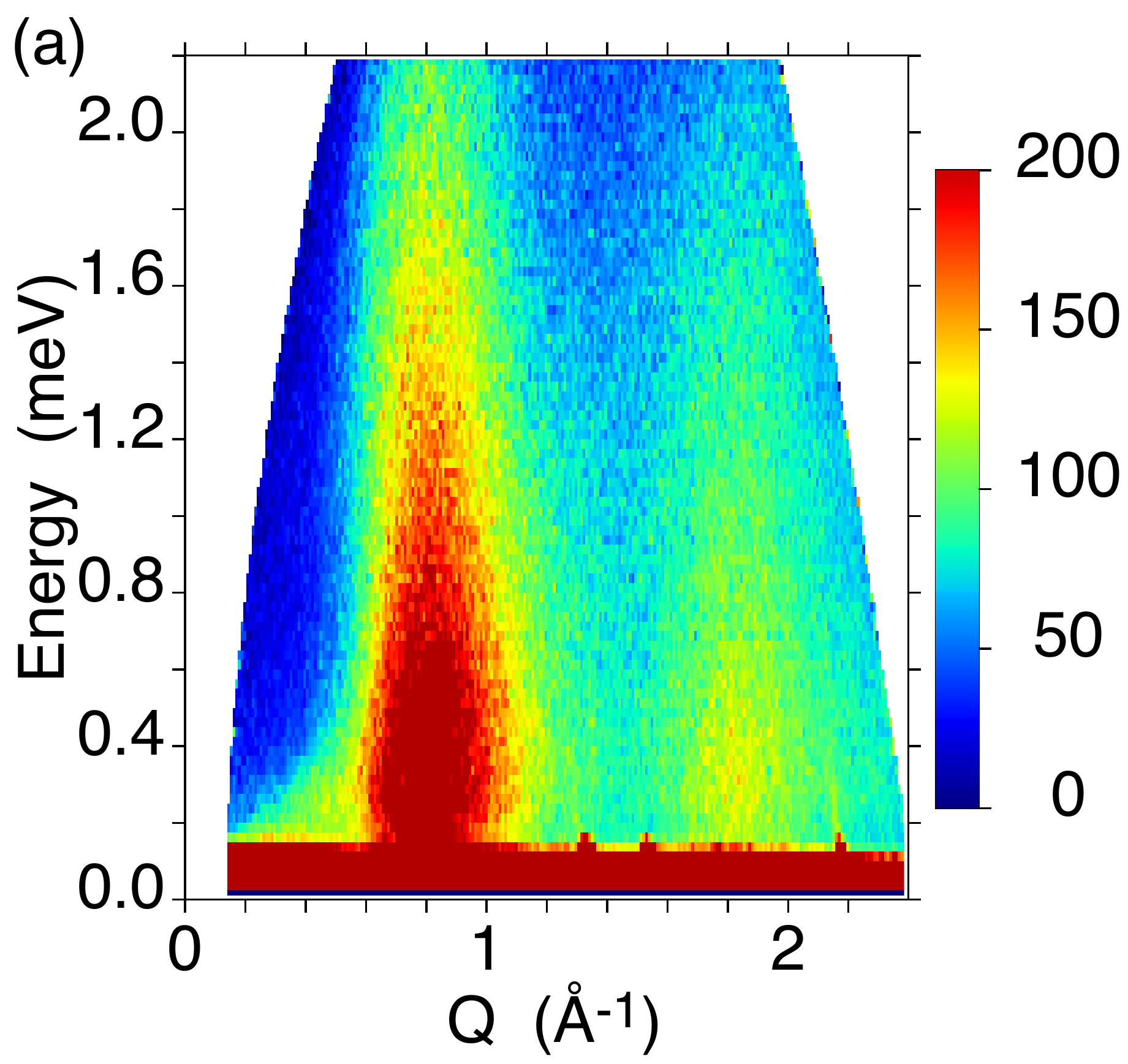}
\includegraphics[width=0.49\columnwidth]{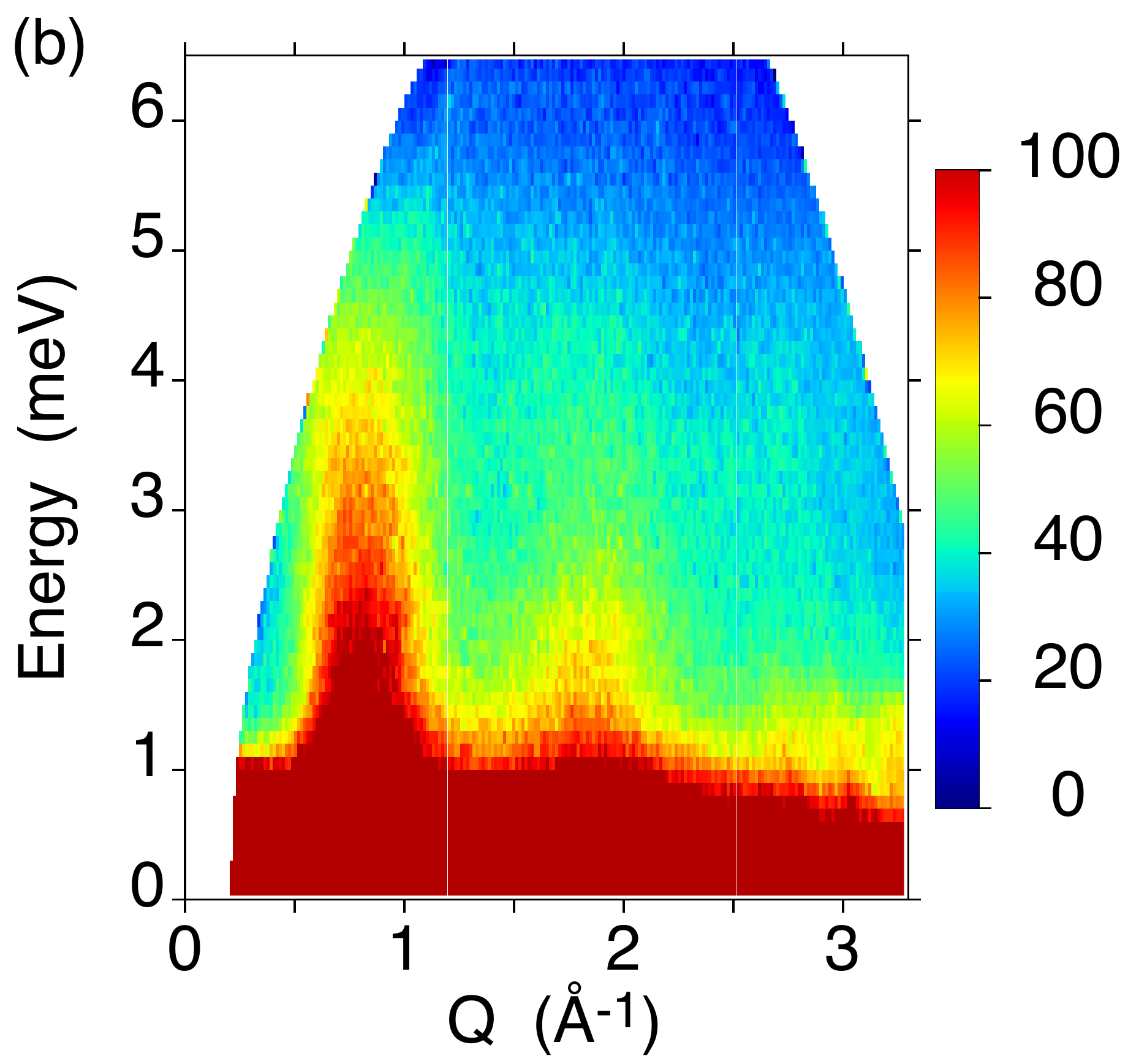}
\includegraphics[width=0.49\columnwidth]{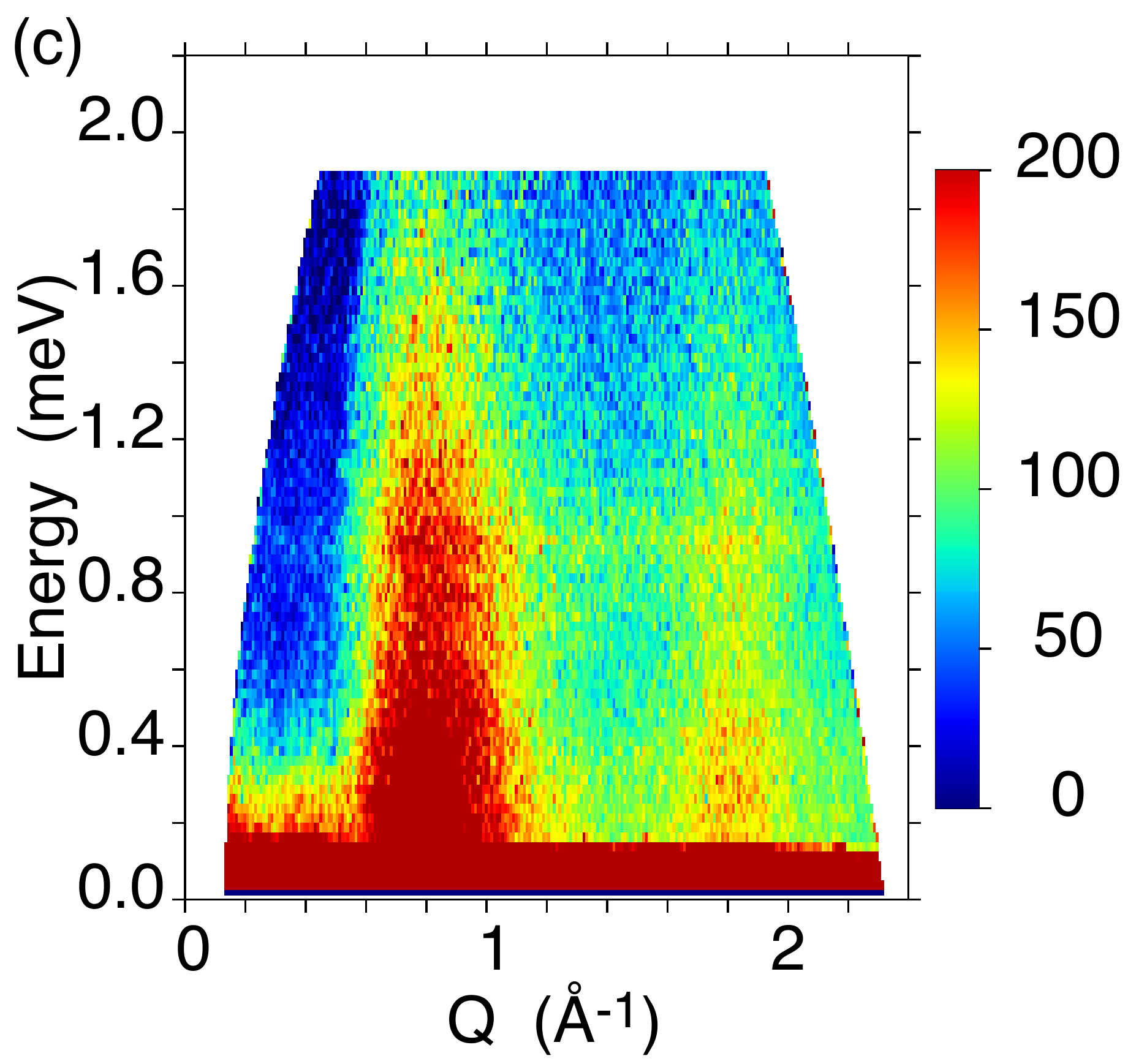}
\includegraphics[width=0.49\columnwidth]{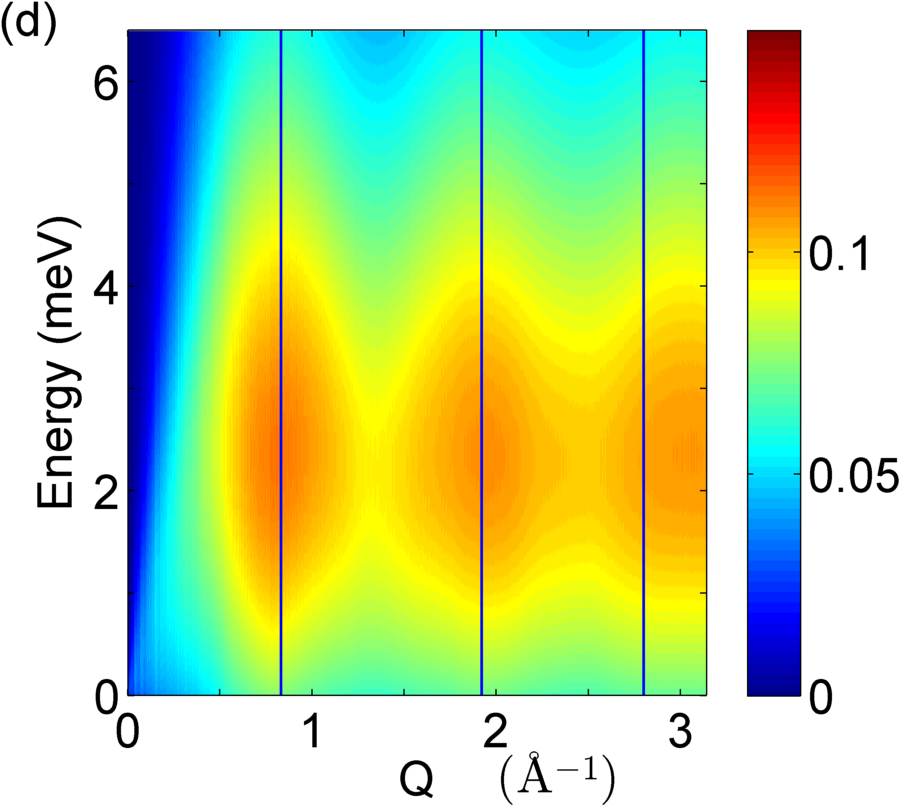}
\caption{
Dynamic susceptibility \XQE\ on a linear intensity scale as a function of wave vector $Q$ and energy $E$ at $T\approx 1.6$~K. (a) \basrnisbo\ with an incoming neutron energy of $E_i=3.55$~meV and (b) with $E_i=8.0$~meV. (c)~\banisbo\ with $E_i=3.27$~meV (the weak feature at 1~meV is an experimental artifact). (d) Calculated powder-averaged \XQE\ of the U(1) Fermi surface state (A) at 1/3 spinon filling (see text).
}
\label{FigMap}
\end{figure}

The powder samples were put in an annular cylinder made from Cu or Al (depending on the temperature range) and thermalized by helium exchange gas. INS measurements were performed on the time-of-flight spectrometer IN5 at the Institut Laue-Langevin, using neutrons with several incident energies $E_i$ between 1.13 and 20.4~meV at temperatures between 0.05 and 150~K using a dilution refrigerator or an orange cryostat. The energy resolution for elastic scattering follows approximately $\Delta E=0.02\times(E_i)^{1.3}$~meV. Standard data reduction \cite{LampRef} including absorption corrections gave the neutron scattering function \SQE, which is related to the imaginary part of the dynamic susceptibility via $\XQE = [1-\exp(-E/k_BT)] \SQE$ shown in Fig.~\ref{FigMap}. The data in this figure are not corrected for the magnetic form factor, and clearly illustrate that scattering from phonons is negligible in the energy and wave-vector range relevant for this work. Our neutron scattering data also show the absence of long-range magnetic order down to $T=0.05$~K in \basrnisbo\ and down to at least $T=1.5$~K in \banisbo.

\begin{figure}[t!]
\includegraphics[width=0.49\columnwidth]{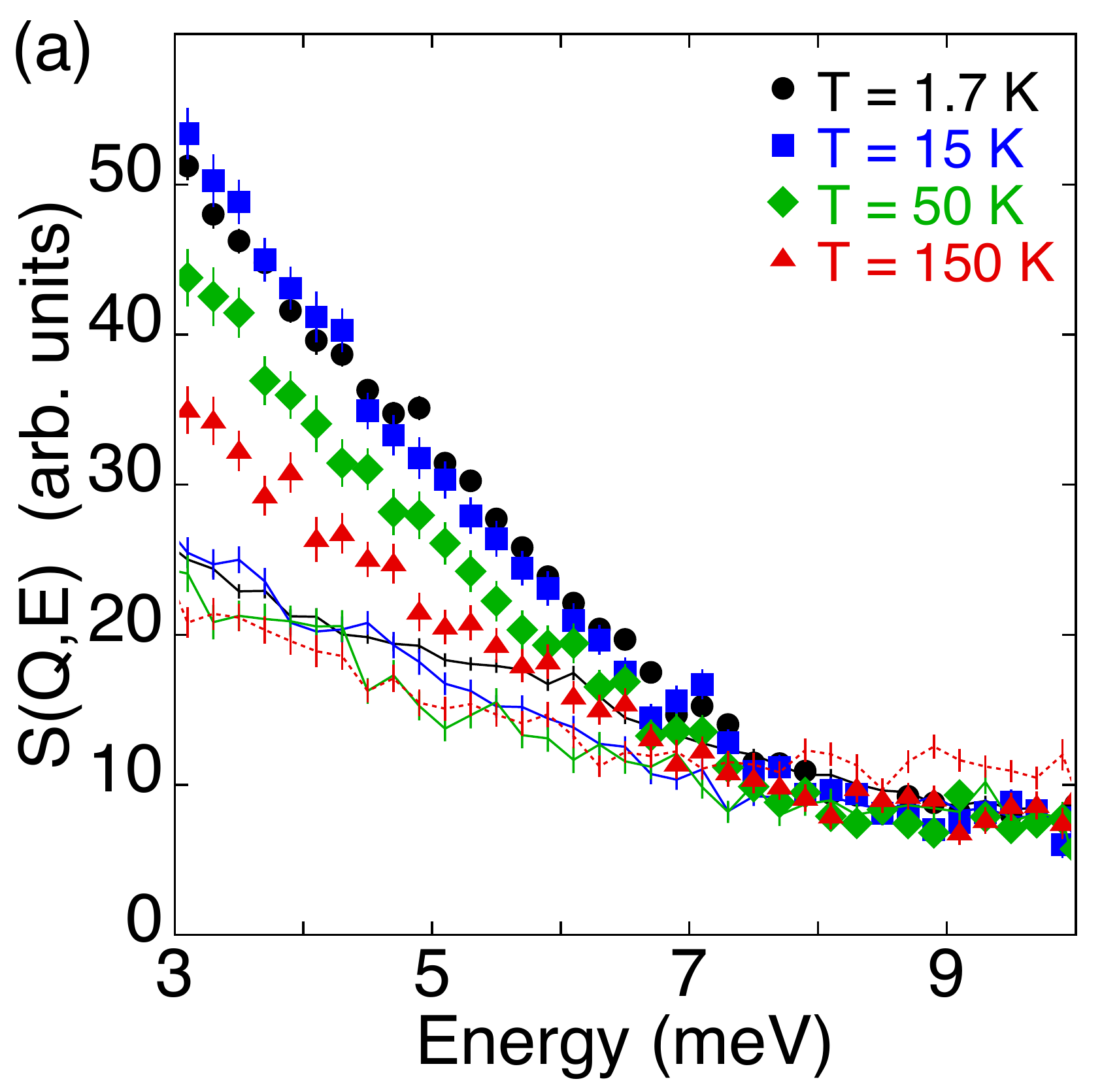}
\includegraphics[width=0.49\columnwidth]{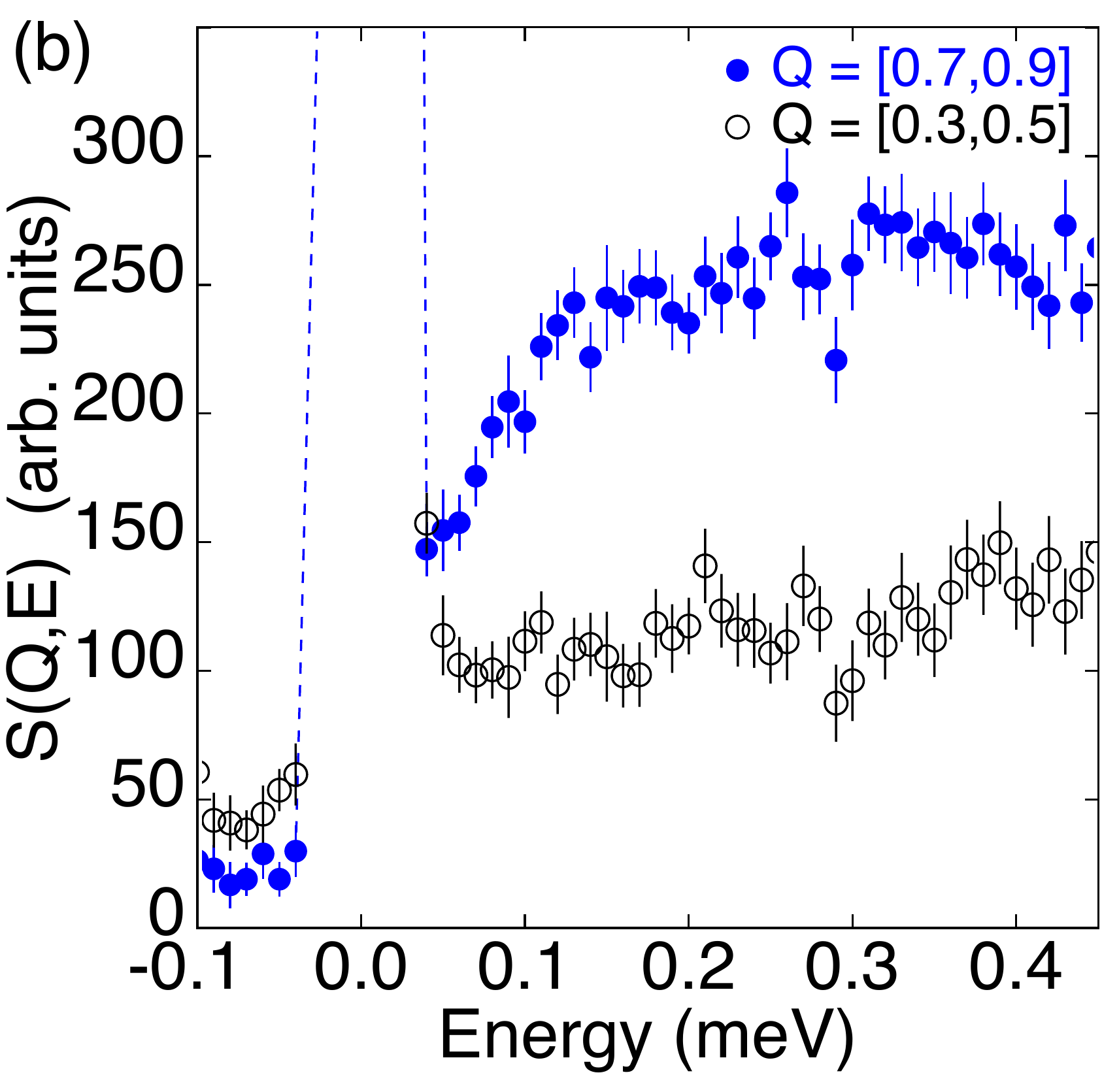}
\includegraphics[width=0.49\columnwidth]{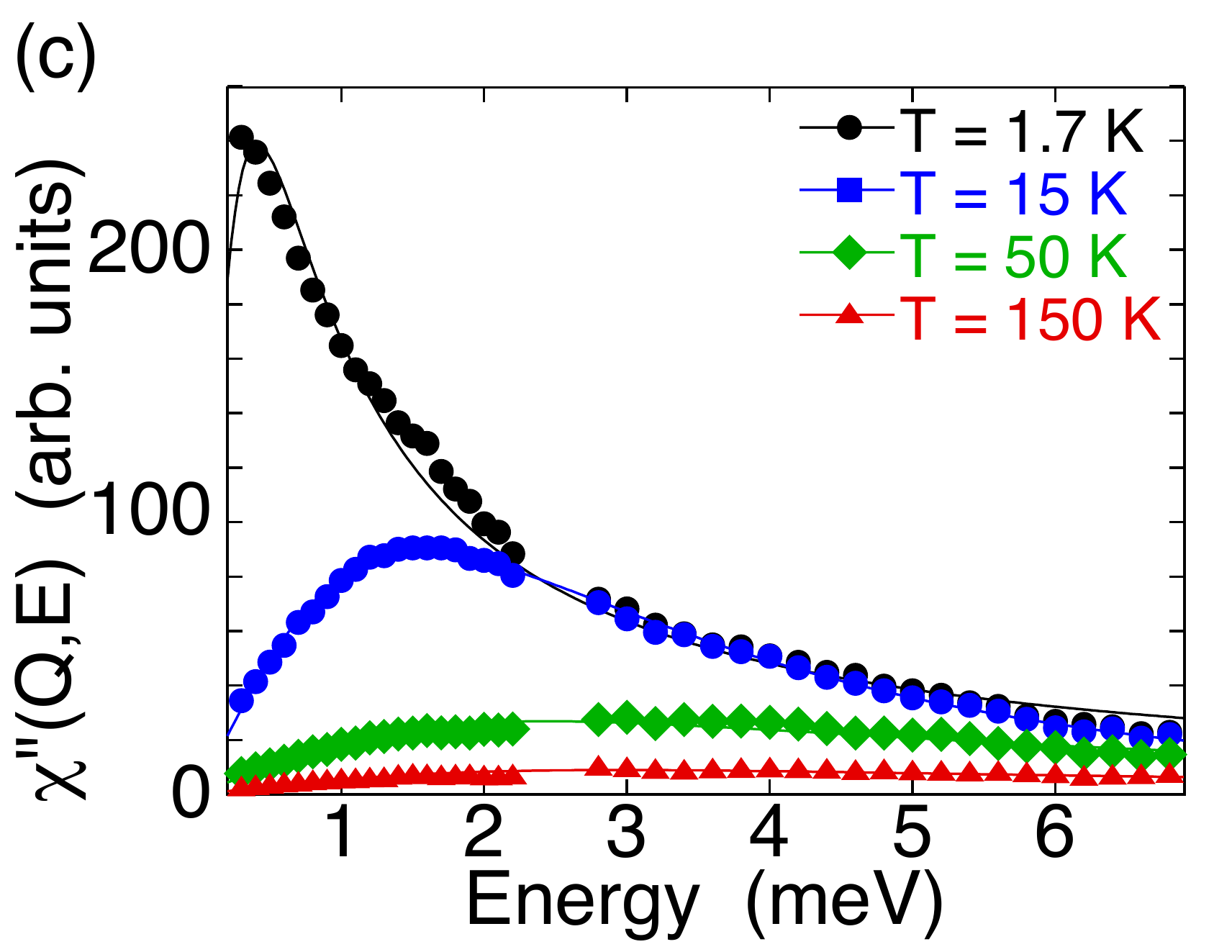}
\includegraphics[width=0.47\columnwidth]{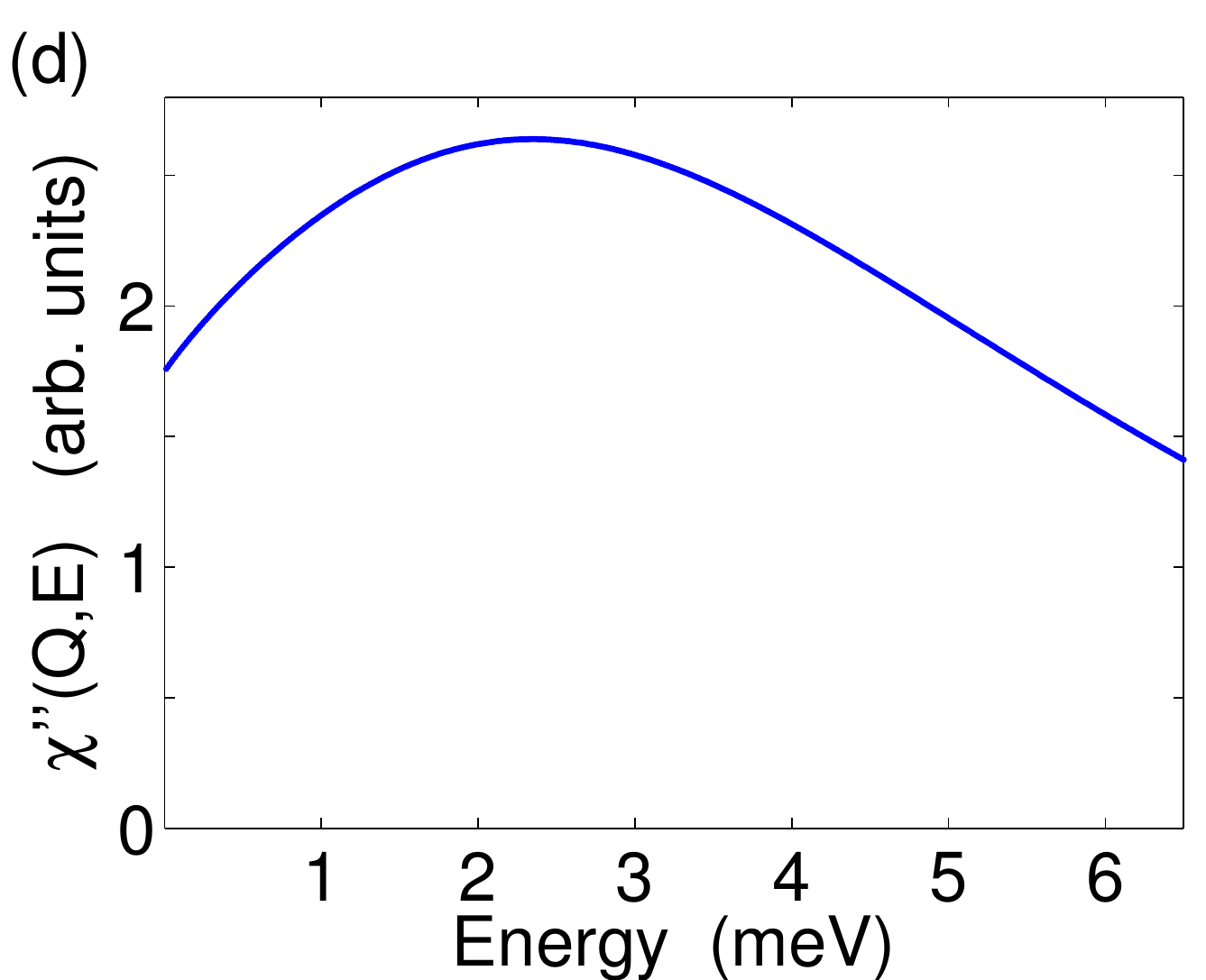}
\caption{
(a)--(c)~Energy dependence of the INS data from \basrnisbo.
Panel (a) shows the temperature dependence of  the dynamic structure factor \SQE\ at wave vectors $Q = 1.9\pm 0.1$~\Ang\ (magnetic signal, symbols) and $Q=3.7 \pm 0.1$~\Ang\ (mostly nonmagnetic signal, lines) measured with an incoming neutron energy of $E_i=14.2$~meV. The excitations extends out to an energy of about 7.5~meV.
Panel (b) shows \SQE\ at $Q = 0.8\pm 0.1$~\Ang\ (magnetic signal, solid blue circles) and $Q=0.4 \pm 0.1$~\Ang\ (mostly nonmagnetic signal, open black circles) for $T=0.1$~K measured with $E_i=1.13$~meV. The excitations are gapless within the experimental energy resolution (the dashed lines indicate the extension of the elastic peak).
Panel (c) shows the imaginary part of the dynamic susceptibility \XQE\ at $Q = 0.8\pm 0.1$~\Ang\ for different temperatures obtained by combining data obtained with $E_i=3.55$ and $14.2$~meV. The lines are guides to the eye.
 (d) Theoretical susceptibility at $Q\simeq 0.8$~\Ang for the U(1) Fermi surface state at $T=0$.
 }
\label{FigEnergy}
\end{figure}

The excitation spectrum shown in Fig.~\ref{FigMap} is characteristic for a spin liquid, with vertical rods of broad scattering coming out at discrete wave vectors. The energy range of these excitations extends out to about 7.5~meV, which can be seen from both the $Q$ and the temperature dependence of \SQE\ [see Fig.~\ref{FigEnergy}(a)]. The intensity extends down to energies below 0.04~meV [see Fig.~\ref{FigEnergy}(b)], i.e., they are gapless within the resolution of the present experiment. This is consistent with the large linear term in the specific heat \cite{Balicas11Ni_PRL.107.197204} and the absence of a gap in NMR measurements of the spin-lattice relaxation rate $1/T_1$ \cite{Quilliam16_PRB.93.214432}. Figure~\ref{FigEnergy}(c) shows the intrinsic energy dependence of the magnetic scattering \XQE\ (i.e., without Bose factor) at $Q=0.8 \pm 0.1$~\Ang\ for different temperatures. The characteristic energy [peak position of \XQE] increases and the intensity decreases with increasing temperature.

\begin{figure}[t!]
\includegraphics[width=0.5\columnwidth]{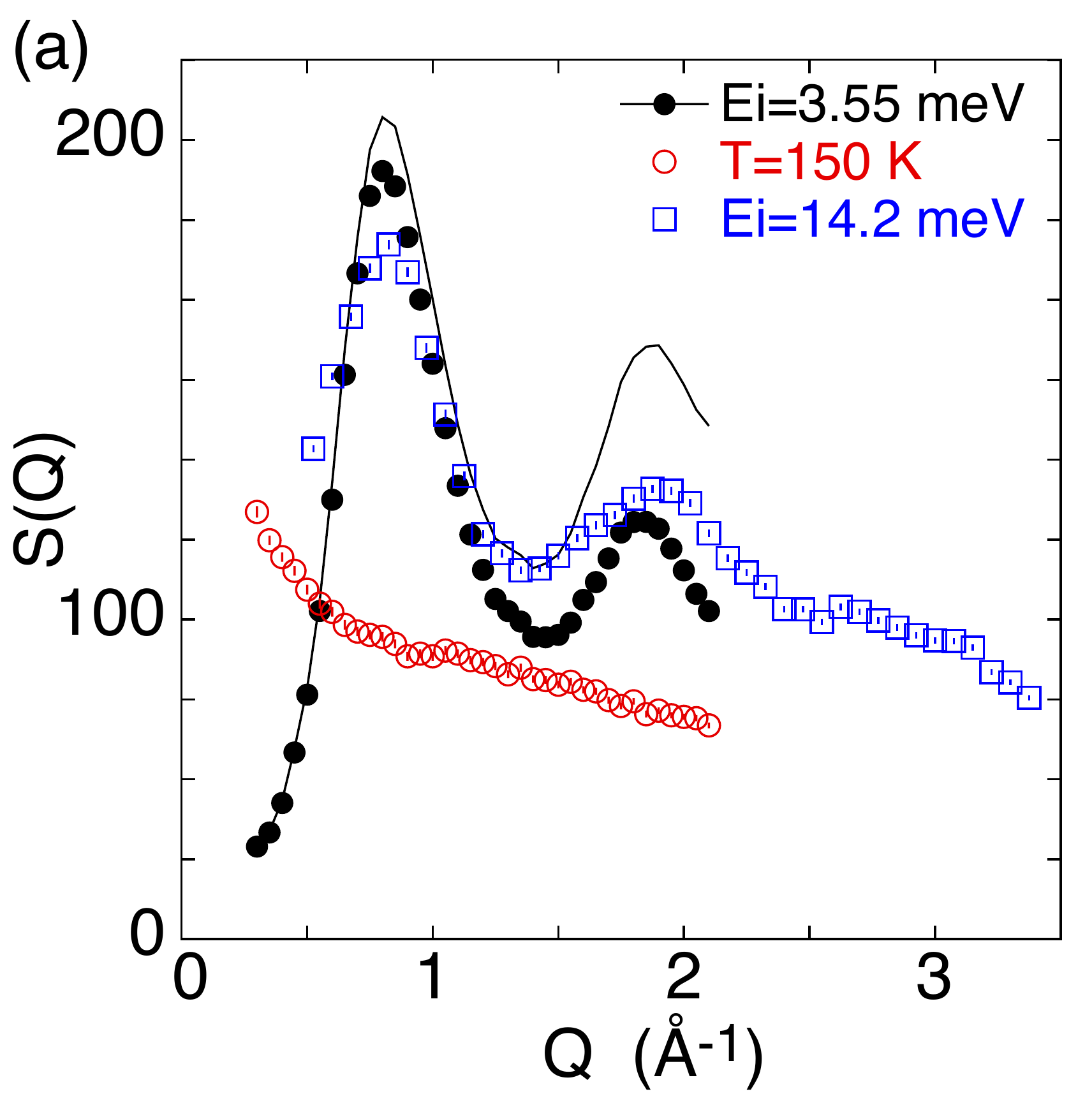}
\includegraphics[width=0.48\columnwidth]{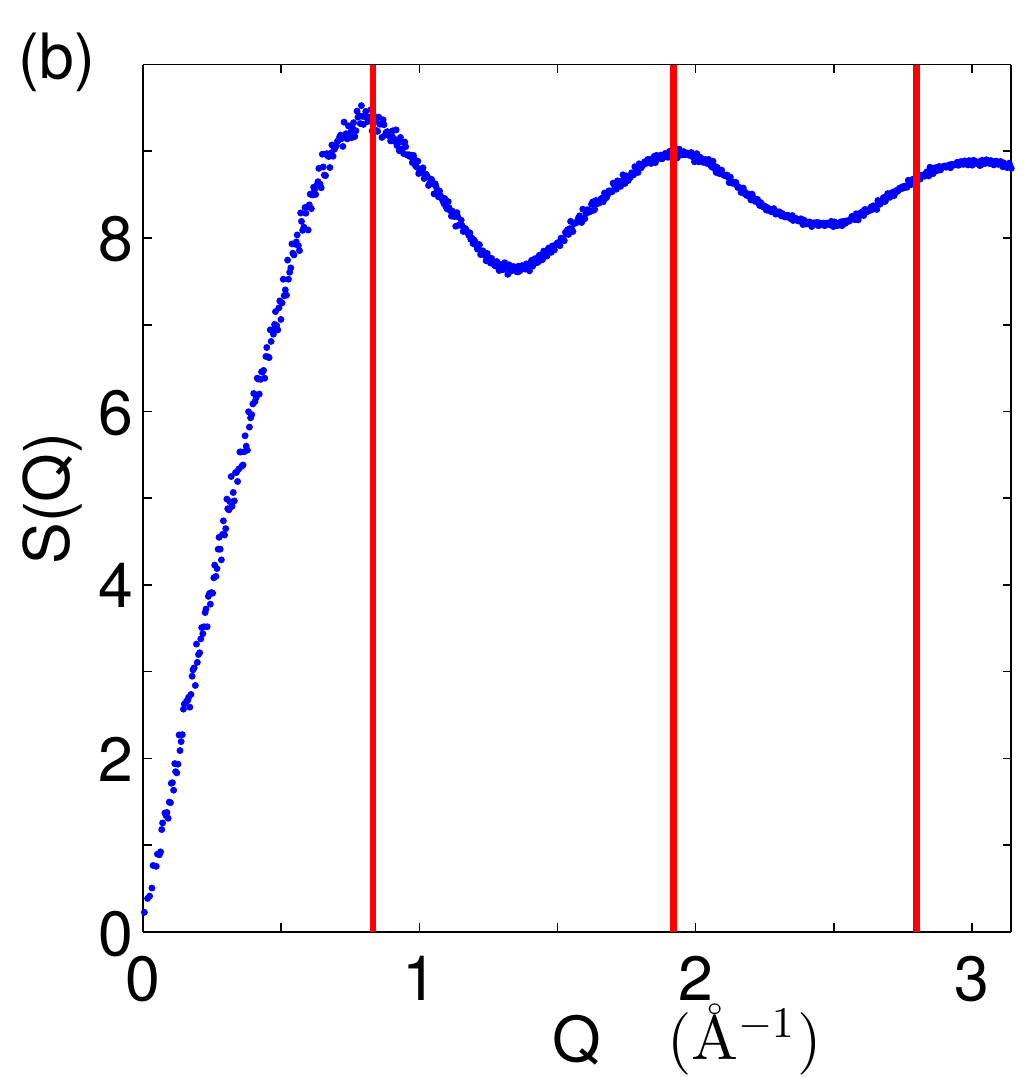}
\caption{
(a) $Q$ dependence of \SQE\ of \basrnisbo\ at $T=1.5$~K integrated over the energy range 0.3--1.5~meV at $E_i=3.55$~meV (solid black circles) and 2--4~meV at $E_i=14.2$~meV (open blue squares) showing peaks at $Q=0.83$ and 1.92~\Ang. The black line shows the $T=1.5$~K data taken at $E_i=3.55$~meV and corrected for the magnetic form factor. The open red circles show data taken with $E_i=3.55$~meV at $T=150$~K, where most of the correlations have disappeared. (b) Calculated static structure factor $S(Q)$ for the U(1) Fermi surface state (A) at $T=0$.
}
\label{FigQ}
\end{figure}

The wave-vector dependence of the scattering after integration over a finite energy interval is shown in Fig.~\ref{FigQ}(a). At low temperatures, the spin-liquid scattering peaks at wave vectors $Q_1=0.83$ and $Q_2=1.92$~\Ang, with a third broad peak at $Q_3=2.8$~\Ang. The data taken with a higher incoming energy is slightly broader due to a wider range for the energy integration. The correlations in $Q$ persist up to at least $T=50$~K (not shown), which confirms the low-dimensional (here, two-dimensional) nature of the magnetic scattering. At even higher temperatures, $T=150$~K, the correlations have almost completely disappeared [red open circles in Fig.~\ref{FigQ}(a)]. The height of the second peak in $S(Q)$ at $Q_2=1.92$~\Ang\ is reduced with respect to the first one, even after correction for the magnetic form factor of the Ni$^{2+}$ ions [see black line in Fig.~\ref{FigQ}(a)]. Attempts to fit the observed structure of $S(Q)$ using broadened Bragg peaks are found to fail. As we will discuss below, the peaks in $S(Q)$ can be attributed to extended regions of reciprocal space, i.e., strong intensity rings close to the boundary of the two-dimensional Brillouin zone~(BZ).

INS measurements were also performed on the pressure synthesized \banisbo\ powder sample. Despite the usage of a high-flux low-resolution configuration with an incoming energy of $E_i = 3.27$~meV, the limited sample quantity led to a strongly reduced statistical quality of the data. Within the precision of these measurements, no major differences were observed compared to the \basrnisbo\ sample [see Fig.~\ref{FigMap}(c)], which makes us confident that the experimental data on the latter are representative of the triangular $6H$-B lattice of \banisbo.

To make further progress, we calculate the static and dynamical spin structure factors for a large number of pertinent gapless quantum spin liquid states for spin $S=1$ on the triangular lattice. For this, we use fractionalization of spin into three~\cite{Liu10_PRB.81.224417} and four~\cite{Xu12_PRL.108.087204} flavors of fermionic spinons. We primarily perform these calculations at the mean-field level, i.e., in the unconstrained Hilbert space, but we have checked that Gutzwiller projection only weakly renormalizes the static susceptibilities and spinon spectra in the relevant cases. More specifically, we investigate three classes of QSLs: (A) the U(1) state with three spinons forming a large Fermi sea discussed in \cite{Bieri12_PRB_86_224409}, (B) the $\mathbb{Z}_4$ QSL with quadratic band touching~(QBT) of four spinons proposed in Ref.~\cite{Xu12_PRL.108.087204}, and (C) a generalization of the recently constructed Dirac spin liquid for triangular spin $S=1/2$ systems~\cite{Bieri16_PRB.93.094437, Lu16_PRB.93.165113, Zheng2015, Iqbal2016} to spin $S=1$ and three spinons, resulting in a state with small spinon Fermi pockets.
Other theoretical proposals for this material are either inconsistent with the gapless and diffuse nature of the measured spin structure factor, and/or are ruled out by the indication of unbroken spin rotation symmetry in recent NMR and $\mu$SR measurements~\cite{Quilliam16_PRB.93.214432}.

The QSL scenarios (A)--(C) have a set of natural parameters that can be related to microscopic spin models. For the Fermi-surface states (A) and (C), we adjust the relative fillings of the three spinons, due to a potential single-ion anisotropy term $D$~\cite{Serbyn11_PRB.84.180403, *Serbyn13_PRB.88.024419, Bieri12_PRB_86_224409}. In the QBT state (B), we add a second-neighbor mean field, related to interaction on that bond.

Among the considered families of states, we find only (A) to be consistent with the INS data. The QSL families (B) and (C) show intensity maxima and minima in their powder-averaged structure factors that are inconsistent with experiment \cite{suppMat}. For state~(A), the agreement is best when all spinons have approximately equal filling of $1/3$, indicating the absence of a sizable $D$ term and unbroken spin-rotation symmetry in the material. The calculated powder-averaged spin structure factor is shown in Fig.~\ref{FigMap}(d). It exhibits broad spinon continua at three wave vectors, extending down to zero energy, consistent with the INS data displayed in the other panels. In Fig.~\ref{FigEnergy}(d), we plot an energy cut of the calculated intensity, integrated over the maximum at $Q = 0.8\pm 0.1$. The strong low-energy weight and ``belly shape'' of this curve are consistent with experiment. However, the intensity is skewed towards high energy, probably due to unphysical components in the mean-field wave function. This may be corrected by invoking Gutzwiller projection removing spinon double occupancies~\cite{Mei15_DSL, DallaPiazza2015, Bieri07_PRB_75_035104}, or by incorporating gauge fluctuations that mediate spinon interaction~\cite{Motrunich05_PRB.72.045105}. Such calculations are beyond the scope of this work \cite{spinonNote}.

We also calculate the bandwidth of Gutzwiller-projected two-spinon excitations in state~(A)~\cite{Bieri15_PRB.92.060407, Hermele08_PRB.77.224413, bwnote}. Assuming a short-range spin model, we estimated $W \simeq 4 J$, where $J$ is the microscopic exchange energy. Using the bandwidth measured in INS, we conclude that $J \simeq 22$~K, in surprisingly good agreement with the observed Curie-Weiss temperature. This energy scale is used in Figs.~\ref{FigMap}(d) and \ref{FigEnergy}(d). For non-interacting spinons, the corresponding hopping amplitude is $t \simeq 16$~K, leading to a large linear term in the specific heat of $\gamma \simeq 0.18 \pi^2 / t \simeq 0.11$~K$^{-1}$ at 1/3 filling of spinons. Furthermore, the Wilson ratio is $R_W = 8/3\simeq 2.7$. These values deviate from the experimental ones ($\gamma^\text{exp} = 0.02$~K$^{-1}$, $R_W^{\text{exp}} = 5.6$ \cite{Balicas11Ni_PRL.107.197204}), which, to some extent, is due to the neglect of spinon interactions in our crude estimates.

\begin{figure}[t!]
\includegraphics[width=0.88\columnwidth]{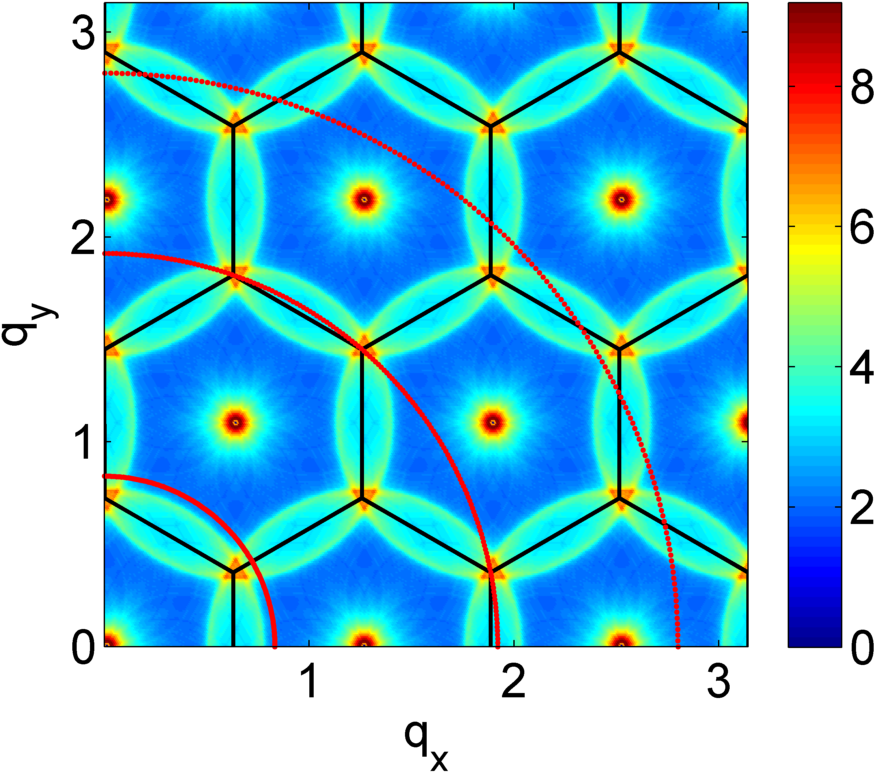}
\caption{
Dynamical susceptibility $\chi''(q, E)$, energy integrated from 0 to 5\% of the spinon bandwidth, in the U(1) Fermi surface state (A) at 1/3 spinon filling. The black hexagons are the BZ boundaries, and the radii of the red circles centered at $\Gamma$ correspond to the three broad intensity maxima seen in the INS data.
}
\label{FigSkw}
\end{figure}

The energy-integrated (static) spin structure factor of state (A) is shown in Fig.~\ref{FigQ}(b), along with the corresponding experimental data in the left panel. The vertical red lines indicate the positions of the measured INS maxima. In Fig.~\ref{FigSkw}, we show the low-energy dynamical spin susceptibility of state (A) in two-dimensional momentum space of the triangular lattice. The BZ boundaries are shown as black hexagons. The continua at the $\Gamma$ points come from $q\sim 0$ two-spinon excitations, while the broad intensities close to the BZ boundary are due to the $q\sim 2 k_F$ excitations of the large spinon Fermi surface. The red circles indicate the $Q$ momenta of the experimental intensity rods seen in INS. The powder average of the intensity in Fig.~\ref{FigSkw}, assuming negligible dispersion in the third direction, is shown in Fig.~\ref{FigMap}(d). This averaging \cite{suppMat} has two crucial effects: First, the broad intensity maxima are slightly shifted to larger $Q$, such that the $2 k_F$ features in Fig.~\ref{FigSkw} accurately reproduce the locations of the INS maxima. Second, the $q\sim 0$ intensities close to $\Gamma$ are washed out and absorbed in a broad background.

The microscopic origin of QSL (A) is not fully understood. The state is known to have a low (but not lowest) variational energy \cite{Bieri12_PRB_86_224409, Serbyn11_PRB.84.180403, *Serbyn13_PRB.88.024419} in a nearest-neighbor Heisenberg model with strong biquadratic interaction \cite{Laeuchli06_PRL.97.087205, Moreno14_PRB.90.144409, Voll15_PRB.91.165128}. A three-site ring exchange term can further stabilize it, but then a triplet pairing sets in, resulting in a chiral $\mathbb{Z}_2$ QSL phase \cite{Bieri12_PRB_86_224409, Serbyn11_PRB.84.180403, *Serbyn13_PRB.88.024419}. Another approach~\cite{Lai13_PRB.87.205131} found that a four-site ring exchange term can stabilize phase (A). We hope our results will stimulate further theoretical work on microscopic mechanisms.

In conclusion, we used inelastic neutron scattering to investigate the magnetic excitation spectrum of powder samples of the $6H$-B phase of \banisbo. Broad gapless and nondispersive excitation continua are observed at characteristic wave vectors. Comparing with several plausible theoretical models, we find that the low-temperature phase realized in this spin $S=1$ Mott insulator is best described by a state of three flavors of unpaired fermionic spinons, and the observed spectrum is consistent with the $2 k_F$ continua of a large two-dimensional spinon Fermi surface.

This work was supported in part by the French Agence Nationale de la Recherche, Grant No.\ ANR-12-BS04-0021. The inelastic neutron scattering experiments were performed at the Institut Laue-Langevin (ILL) in Grenoble, with doi:10.5291/ILL-DATA.4-01-1401. SB acknowledges the hospitality and support of the ILL and the International Institute of Physics at the Universidade Federal do Rio Grande do Norte, Natal, Brazil. We thank B.~Bernu, F.~Bert, M.~Enderle, P.~A.\ Lee, C.~Lhuillier, and J.~A.\ Quilliam for helpful discussions.

\bibliography{trg_fs_biblio}

\begin{thebibliography}{51}
\bibitem [{spi()}]{spinonNote}%
  \BibitemOpen
\bibinfo {note} {In the calculated dynamical susceptibilities
  [Figs.~\ref{FigMap}(d), \ref{FigEnergy}(d), and \ref{FigSkw}], we use a
  spinon lifetime broadening $\Gamma \sim |\omega|^{2/3}$, as suggested by
  coupling to a U(1) gauge field~\cite{Lee09_PRB.80.165102}. This broadens the
  intensity and moves it down in energy, but it does not affect the main
  conclusions of this work.}\BibitemShut {Stop}%
\bibitem [{bwn()}]{bwnote}%
  \BibitemOpen
\bibinfo {note} {Theoretically, we define the bandwidth by taking a spinon from
  the Fermi surface to an outermost level. Spin excitations constructed taking
  spinon states from far below to far above the Fermi energy are expected to be
  strongly suppressed. See also \cite{suppMat}.}\BibitemShut {Stop}%
\end{thebibliography}



\section*{Supplementary material}

# Supplemental Material: Evidence for a spinon Fermi surface in the triangular $S = 1$ quantum spin liquid $Ba_3NiSb_2O_9$

B. Fåk,[1,][*] S. Bieri,[2,3,][†] E. Canévet,[1,4,5] L. Messio,[3] C. Payen,[6]
M. Viaud,[6] C. Guillot-Deudon,[6] C. Darie,[7] J. Ollivier,[1] and P. Mendels[8]

[1]*Institut Laue-Langevin, CS 20156, 38042 Grenoble Cedex 9, France*
[2]*Institute for Theoretical Physics, ETH Zürich, 8099 Zürich, Switzerland*
[3]*Laboratoire de Physique Théorique de la Matière Condensée,*
*CNRS UMR 7600, Université Pierre et Marie Curie,*
*Sorbonne Universités, 75252 Paris, France*
[4]*Laboratory for Neutron Scattering & Imaging,*
*Paul Scherrer Institut, 5232 Villigen, Switzerland*
[5]*Laboratoire de Physique des Solides, Université Paris Sud 11, CNRS UMR 8502, 91405 Orsay, France*
[6]*Institut des Matériaux Jean Rouxel, CNRS UMR 6502,*
*Université de Nantes, 44322 Nantes Cedex 3, France*
[7]*Institut Néel, CNRS, Univ. Grenoble Alpes, BP166, 38042 Grenoble Cedex, France*
[8]*Laboratoire de Physique des Solides, CNRS, Université Paris-Sud,*
*Université Paris-Saclay, 91405 Orsay Cedex, France*

(Dated: December 22, 2016)

To supplement the main text of the paper, we provide here additional information on the synthesis of the powder sample of $Ba_{2.5}Sr_{0.5}NiSb_2O_9$ and its characterization by X-ray diffraction and bulk magnetization. We also provide theoretical results concerning (i) the spin structure factors for some alternative plausible spin-1 QSL states; (ii) spinon mean-field spectrum for QSL state (A) of the main text; (iii) effects of Gutzwiller projection on the static spin susceptibility of state (A); (iv) the formula used to calculate powder averages; and (v) energies of Gutzwiller-projected two-spinon excitations in the U(1) QSL (A) with large Fermi surface.

## A. Synthesis and characterization of $Ba_{2.5}Sr_{0.5}NiSb_2O_9$

### 1. Synthesis

A 7-g powder sample of $Ba_{2.5}Sr_{0.5}NiSb_2O_9$ was prepared by heating a stoichiometric pelletized mixture of high-purity barium carbonate ($BaCO_3$), strontium carbonate ($SrCO_3$), antimony(V) oxide ($Sb_2O_5$), and nickel oxide (NiO) at 1200–1350°C for several days in air with several intermediate grindings. The sample was furnace cooled at the end of the final heat treatment.

### 2. X-ray powder diffraction

An X-ray diffraction (XRD) pattern was collected at room temperature on a Bruker D8 Advance instrument using monochromatic $Cu_{K-L3}$ ($\lambda = 1.540598$ Å) X-rays and a LynxEye detector. Rietveld analyses of the XRD data were performed using JANA 2006.[1] The XRD pattern showed very narrow diffraction peaks and no sign of unreacted starting materials. Table I shows the refined structural parameters and final agreement factors obtained using the published $P6_3/mmc$ crystal structure of the 6H-B phase of $Ba_3NiSb_2O_9$ [2] as a starting model. The composition of the sample was constrained to be $Ba_{2.5}Sr_{0.5}NiSb_2O_9$. The Sr atoms were equally distributed over the two Ba sites. The displacement parameters of the two crystallographically distinct oxygen atoms were constrained to be equal. The atomic coordinates listed in Table I are in good agreement with those

TABLE I. Refined structural parameters and final agreement factors for $Ba_{2.5}Sr_{0.5}NiSb_2O_9$ derived from X-ray diffraction data collected at $T = 300$ K. Space Group $P6_3/mmc$ (No. 194): $a = 5.7728(2)$ Å, $c = 14.2430(3)$ Å, $\chi^2 = 1.27$, $R_p = 10.75$, $R_{wp} = 16.37$.

| Atom | Site | $x$ | $y$ | $z$ | $U_{iso}$ (Å$^2$) | Occupancy |
|---|---|---|---|---|---|---|
| Ba1/Sr1 | 2b | 0 | 0 | 0.25 | 0.007(1) | Ba: 0.83 |
| | | | | | | Sr: 0.17 |
| Ba2/Sr2 | 4f | 1/3 | 2/3 | 0.1000(2) | 0.013(1) | Ba: 0.83 |
| | | | | | | Sr: 0.17 |
| Sb1 | 2a | 0 | 0 | 0 | 0.003(1) | 1 |
| Sb2/Ni1 | 4f | 1/3 | 2/3 | 0.6570(2) | 0.006(1) | Ni: 0.50 |
| | | | | | | Sb: 0.50 |
| O1 | 6h | 0.508(2) | 0.016(4) | 0.25 | 0.013(3) | 1 |
| O2 | 12k | 0.165(2) | 0.330(4) | 0.5800(9) | 0.013(3) | 1 |

determined for the 6H-B phase of $Ba_3NiSb_2O_9$.[2] The refined lattice parameters, $a = 5.7728(2)$ and $c = 14.2430(3)$ Å, were found to be slightly smaller than those observed for the 6H-B phases of $Ba_3NiSb_2O_9$.[2,3] Figure 1 shows the corresponding final Rietveld plot.

Based on the results reported in Ref. 2, the refinement shown in Table I and Fig. 1 corresponds to an averaged structure made of two types of domains with different stacking sequences of the Ni and Sb atoms on the $4f$ site that form the face-sharing $NiSbO_9$ bi-octahedra; Sb-Sb=Ni-Sb and Sb-Ni=Sb-Ni along the $c$ axis, where the "=" represents face-sharing. From XRD, the two stacking sequences appear as equiprobable and are randomly distributed along the stacking axis ($c$ axis). On a local scale, each domain consists of extended regions of Ni atoms forming triangular layers. This conclusion is reinforced by the magnetic susceptibility measurements discussed next.

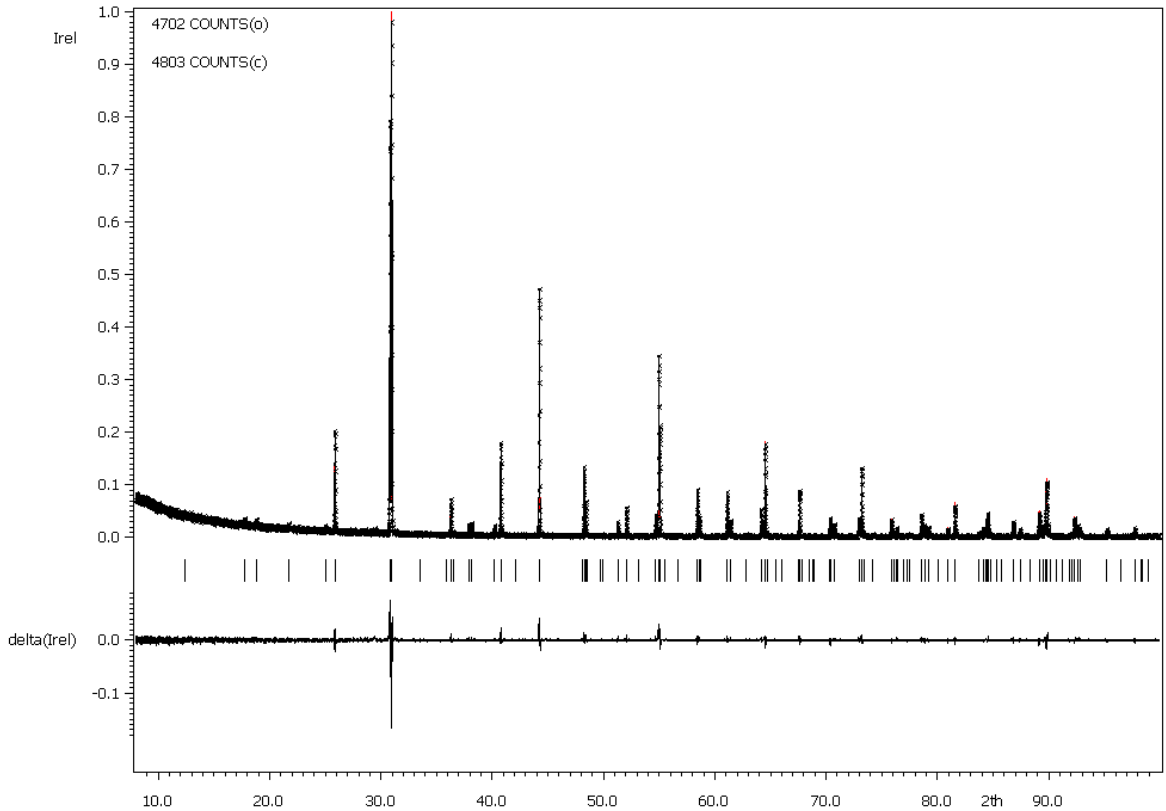

FIG. 1. Observed and calculated X-ray powder diffraction pattern of the $Ba_{2.5}Sr_{0.5}NiSb_2O_9$ sample used for the inelastic neutron scattering experiments. The vertical ticks indicate the positions of the Bragg reflections. The lower curve shows the difference between the observed and calculated data on the same scale.

### 3. Bulk magnetic susceptibility

A commercial SQUID magnetometer (Quantum Design, MPMS-XL7) was used to collect DC magnetization data from $T = 2$ to $300$ K in an applied field of $\mu_0 H = 0.5$ T. Data were corrected for the diamagnetism of the sample holder as well as for core diamagnetism using Pascal's constants.[4] Figure 2 shows the temperature dependent magnetic susceptibility. Data obtained for our sample of pure 6H-B $Ba_3NiSb_2O_9$ is also shown for the sake of comparison. Both samples show a Curie-like tail at low-temperature which correspond to about 2% of $S = 1$ orphan spins.

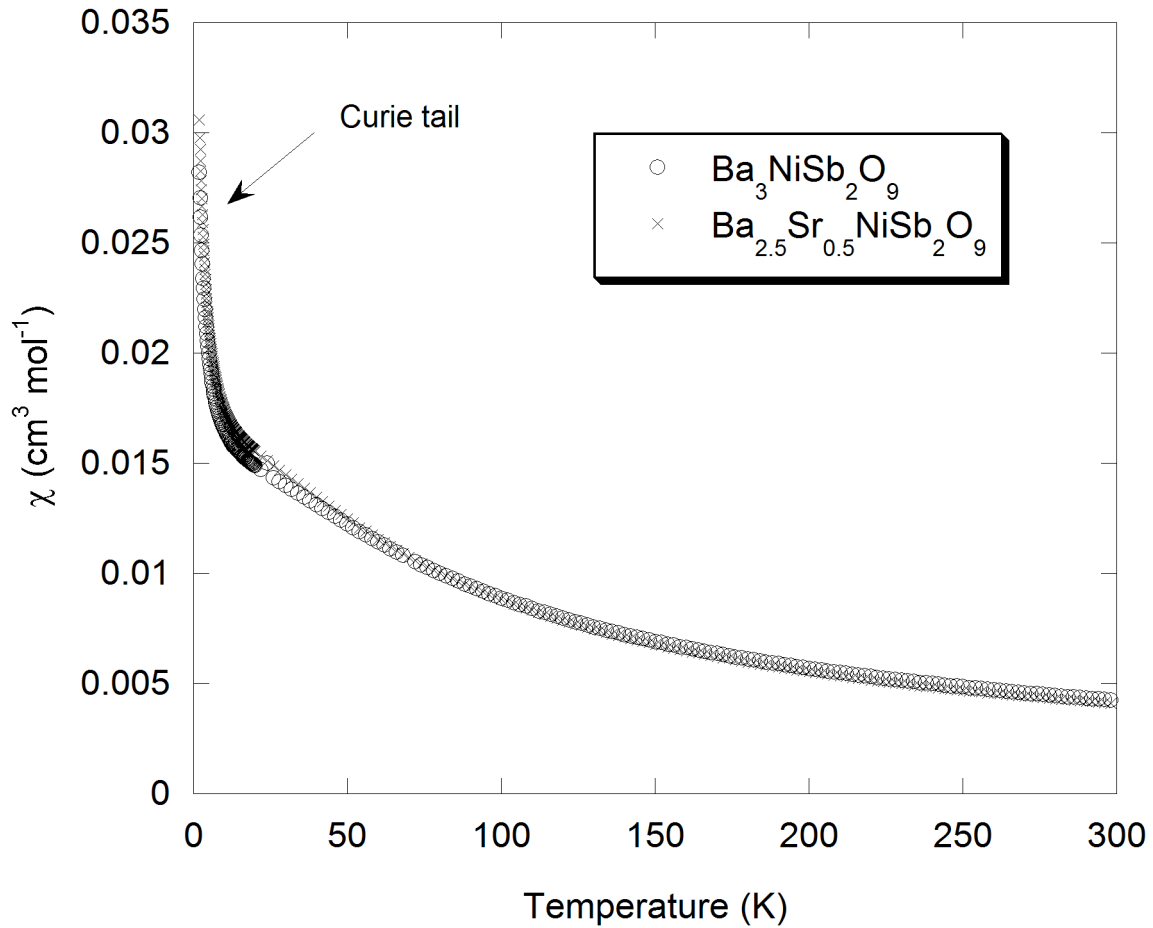

FIG. 2. Magnetic susceptibility $\chi$ versus temperature for our polycrystalline samples of $Ba_{2.5}Sr_{0.5}NiSb_2O_9$ and 6H-B $Ba_3NiSb_2O_9$.

## B. Spin structure factors for alternative spin-1 QSL states

In this section, we present the static and dynamic spin susceptibilities of some pertinent quantum spin liquid states of fractionalized spin $S = 1$. In all dynamical quantities presented here, we use sharp spinons (lifetime broadening $\Gamma \lesssim 1\%$ of the spinon bandwidth) for simplicity. The temperature is set to zero.

In Fig. 3, we present properties of state (A) discussed in the main text, i.e., three flavors of spinons hopping on the triangular lattice, forming a large Fermi surface at $1/3$ filling.[5,6] In the left panel, we show the powder-averaged static structure factor $S(Q)$ in black, and the low-energy intensities $S(Q,\omega)$ in blue. The dynamical quantities are energy-integrated from 0 to 5% of the spinon bandwidth, from 5% to 10%, etc. The lowest blue curve is the lowest energy, etc. These curves may be compared with Fig. 1(d) of the main text. (Note that a stronger lifetime broadening is used in the main text, washing out the intensity at small $Q$.)

In Fig. 4, we show case (C) of the main text,[7–10] i.e., the "Dirac QSL" at filling $1/3$, where spin-rotation symmetry is unbroken and the spinons form Fermi pockets. This structure with small circular $k_F$ features can be well appreciated in the low-energy intensity shown in the middle panel. From the powder-averaged structure factor in the left panel, it is evident that the inelastic intensity is inconsistent with such a QSL state.

In Fig. 5, we show the same state as above, but at filling $1/2$, where two of the three spinons have their chemical potentials at the Dirac points of the spectrum while the third spinon is unoccupied. (In contrast to the previous case, this may now be literally called a "Dirac QSL". It is related to the spin $S = 1/2$ Dirac QSL discussed in Refs. 7–11.) Again, the presence of gapless Dirac points at the Fermi energy in the spinon spectrum is well visible in the low-energy intensity shown in the middle panel. This state breaks spin rotation symmetry, and the measured inelastic intensity spectra are inconsistent, as best seen from the left panel.

In Fig. 6, we display the structure factors for a spin fractionalization with *four* spinons, forming a $\mathbb{Z}_4$ state with quadratic bands touching at the Fermi level [case (B) discussed in the main text].[12] One can see that the intensity in the powder-averaged structure factor at low energy is inconsistent with inelastic intensity data. Here, we show only the case of a first-neighbor spinon mean field, but the conclusion is the same when a second-neighbor parameter is introduced.

Note that the powder-averaged static spin structure factors (black curves in the left panels of Figs. 3 – 6) are rather similar for all states we display here. This is due to the fast decaying (algebraically) spin-spin correlations in all these liquids, leading to very broad intensity maxima in

$S(\boldsymbol{q})$ at the corners of the Brillouin zone (i.e., at the $K$ points). Hence, this comparison with the data does not allow an unambiguous identification. The states, however, show clearly distinctive features in their low-energy intensities, allowing for a unique identification of the best candidate wave function from the inelastic intensity data.

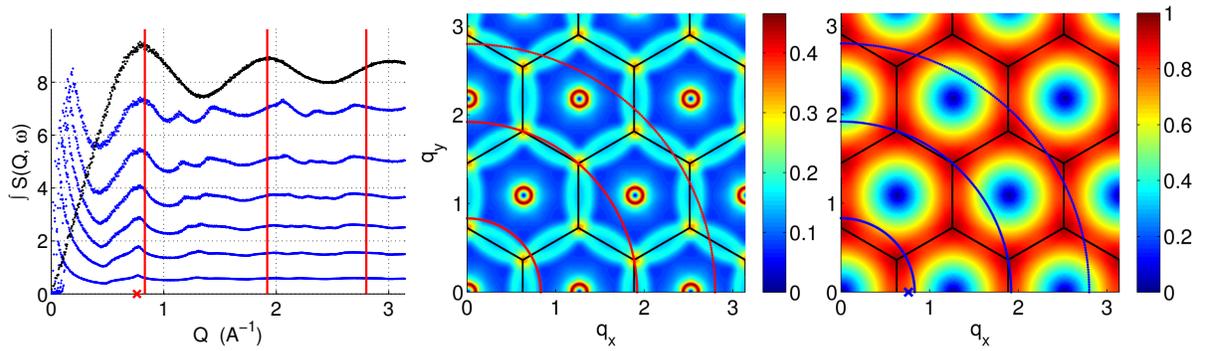

FIG. 3. U(1) QSL state with large spinon Fermi surface at filling 1/3; case (A) in the main text. Left: Powder-averaged static (black) and low-energy (blue) spin susceptibility. Red lines are locations of experimental peaks. Middle: Low-energy intensity, $S(\boldsymbol{q}, \omega)$ integrated from $\omega = 0$ to 5% of the spinon bandwidth; Right: Static structure factor $S(\boldsymbol{q})$. The circles are experimental peak locations. All quantities are calculated at zero temperature in the mean-field wave function.

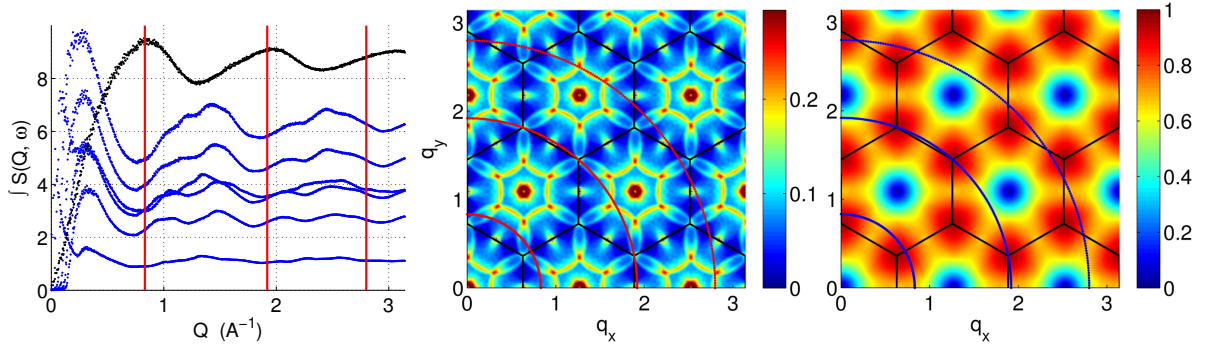

FIG. 4. U(1) QSL state with spinon Fermi pockets; case (C) at filling 1/3. Quantities shown as in Fig. 3.

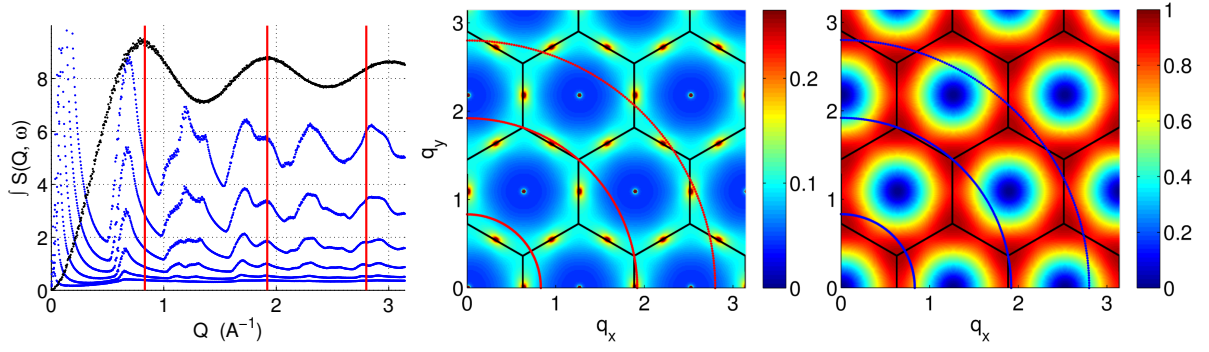

FIG. 5. U(1) QSL state with Dirac spectrum; case (C) at filling 1/2. Quantities shown as in Fig. 3.

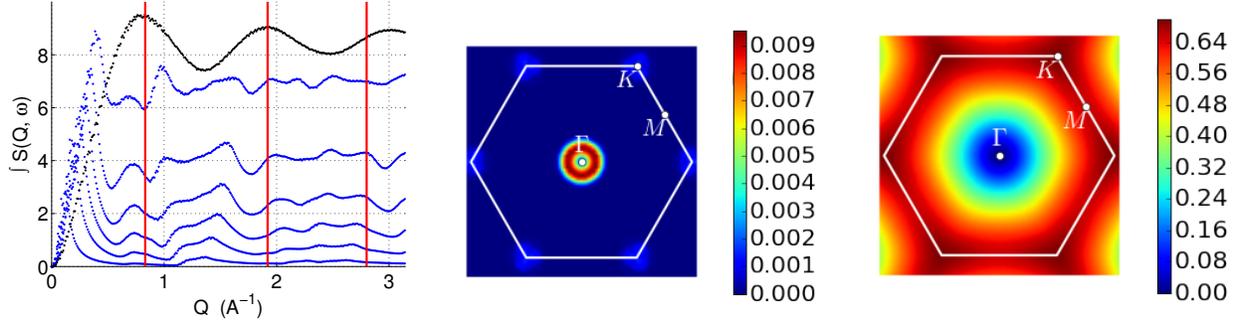

FIG. 6. $\mathbb{Z}_4$ QSL quadratic spinon band touching; case (B) in the main text. Quantities shown as in Fig. 3. Second-neighbor mean field $\Delta_2 = 0$.

## C. Spinon mean-field spectrum

In Fig. 7, we show the mean field spectrum for QSL (A) at filling $1/3$.[5,6] The first Brillouin zone (BZ) is shown in blue. The red dashed line is the large spinon Fermi surface. The states below the Fermi energy ($\epsilon_F = 0$) are occupied, while those above are empty in the ground state. One can see that the Fermi surface is almost circular, leading to circular $2k_F$ features in Fig. 3 (middle panel). It is interesting to note that $k_F \simeq 0.65\pi/a$, which is about 49% of the distance $\overline{\Gamma K}$. This leads to almost commensurate $2k_F$ features at the $K$ points in Fig. 3.

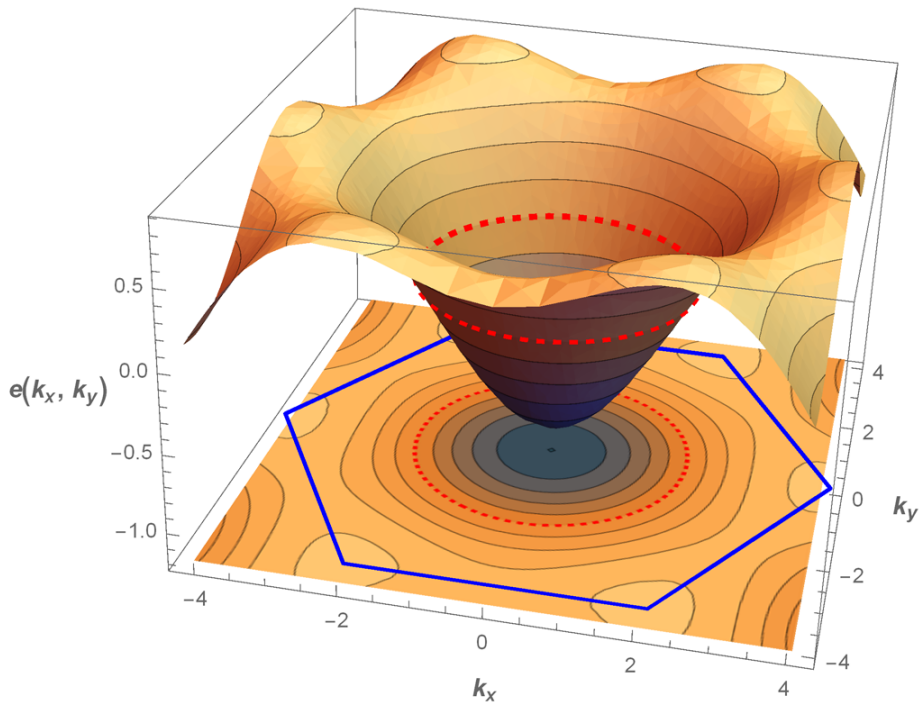

FIG. 7. Spinon mean field spectrum for QSL state (A) at filling $1/3$. The spinon Fermi surface is the red dashed line, the first Brillouin zone is shown in blue.

## D. Gutzwiller projected spin structure factor

In Fig. 8, we illustrate the effect of Gutzwiller projection on the static spin structure factor $S(\bm{q})$ for state (A) of the main text. The Gutzwiller projection is done on a square cluster of $18 \times 18$ sites. Comparing the middle panel (mean field) and the right panel (projected), we see that the broad peaks at the $K$ points of the BZ are sharpened by the Gutzwiller projection. However, the overall properties of the structure factor remain intact. This is reflected in the left panel, where we show the corresponding powder averages: The peak locations are unchanged by projection, but the width becomes slightly smaller. Also, the peaks at larger wave vector $Q$ are slightly reduced in intensity with respect to the first peak. It is plausible that Gutzwiller projection changes the power law[13] of the spin-spin correlation in real space of this algebraic spin liquid, similar to the situation in one dimension. However, due to limited system size, we have refrained from a detailed analysis of long-distance properties here.

Calculating the effect of projection on the *dynamical* properties of spectral functions in fractionalized quantum spin liquids is beyond the scope of this work. See the following references for recent progress.[14–18] However, sharp features in the spectral function, e.g., due to $2k_F$ excitations, are expected to survive projection.

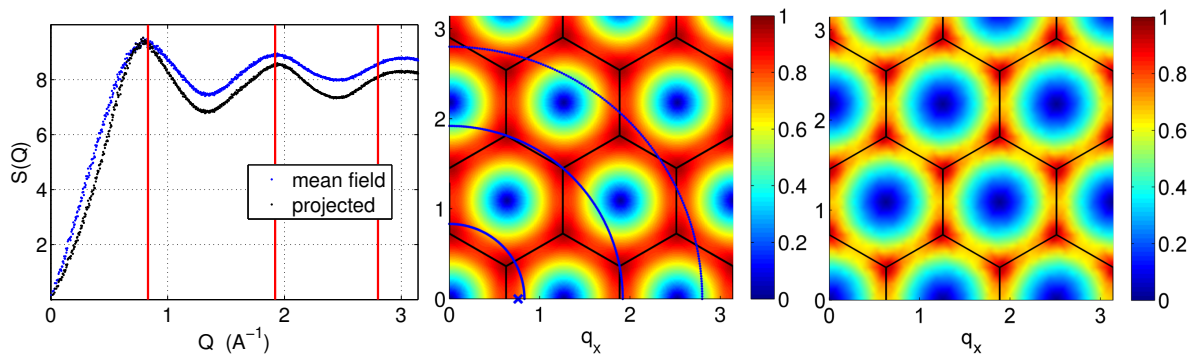

FIG. 8. Effect of Gutzwiller projection on the static spin structure factor of U(1) state with large Fermi surface at 1/3 filling [case (A) in the main text]. Left: Powder-averaged structure factors. Middle: $S(\bm{q})$ before projection (mean field). Right: $S(\bm{q})$ after projection, $18 \times 18$ sites.

### E. Powder average

Here, we denote the three-dimensional momentum by $\boldsymbol{Q} = (q_x, q_y, q_z)$, where $\boldsymbol{q} = (q_x, q_y)$ lies in the two-dimensional layers of the sample, and $q_z$ is normal to the layers. The intensity measured in inelastic neutron scattering on a poly-crystallin sample is proportional to the powder-averaged spin structure factor $S(Q, \omega)$. It is given by the structure factor of the three-dimensional single-crystal sample, $S(\boldsymbol{Q}, \omega)$, averaged over all momentum directions, keeping its norm fixed at $|\boldsymbol{Q}| = Q$. This can be written as

$$S(Q, \omega) = \int S(\boldsymbol{Q}, \omega) \sin\theta \, d\theta \, d\varphi \,, \tag{1}$$

where $\boldsymbol{Q} = (\sin\theta \cos\varphi, \sin\theta \sin\varphi, \cos\theta)Q$.

Next, we assume that the sample is two dimensional, i.e., the dependency of $S(\boldsymbol{Q}, \omega)$ on $q_z$ can be neglected. In this case, we have $S(\boldsymbol{Q}, \omega) = S(\boldsymbol{q}, \omega)$, and the powder average is

$$S(Q, \omega) = \int S(\boldsymbol{q}, \omega) \sin\theta \, d\theta \, d\varphi \,, \tag{2}$$

with $\boldsymbol{q} = (\sin\theta \cos\varphi, \sin\theta \sin\varphi)Q$, and $S(\boldsymbol{q}, \omega)$ is the structure factor of a the two-dimensional state. Substituting $q = Q \sin\theta$ in the integral, we obtain

$$S(Q, \omega) = \int_0^Q S(\boldsymbol{q}, \omega) \frac{q \, dq \, d\varphi}{Q\sqrt{Q^2 - q^2}} \,, \tag{3}$$

with $\boldsymbol{q} = (\cos\varphi, \sin\varphi)q$. This formula is used in this Supplemental Material and in the main part of this paper to calculate the powder average of the two-dimensional structure factor.

The powder average, Eq. (3), has the following mathematical properties. If the two-dimensional structure factor exhibits sharp (delta-)peaks at certain $\boldsymbol{q}$ locations (e.g., due to magnon excitations and long-range order), these peaks remain at the same locations in the average, $Q^{\max} = |\boldsymbol{q}^{\max}|$, but the intensity is slightly smeared to larger $Q$, see, e.g., Ref. [19]. On the other hand, in the case of very broad intensities in $S(\boldsymbol{q}, \omega)$ as encountered in quantum spin liquid phases, the powder average (3) can shift the maxima to larger $Q$, $Q^{\max} \gtrsim |\boldsymbol{q}^{\max}|$. This is what happens, e.g., in the U(1) Fermi surface QSL in Fig. 3.

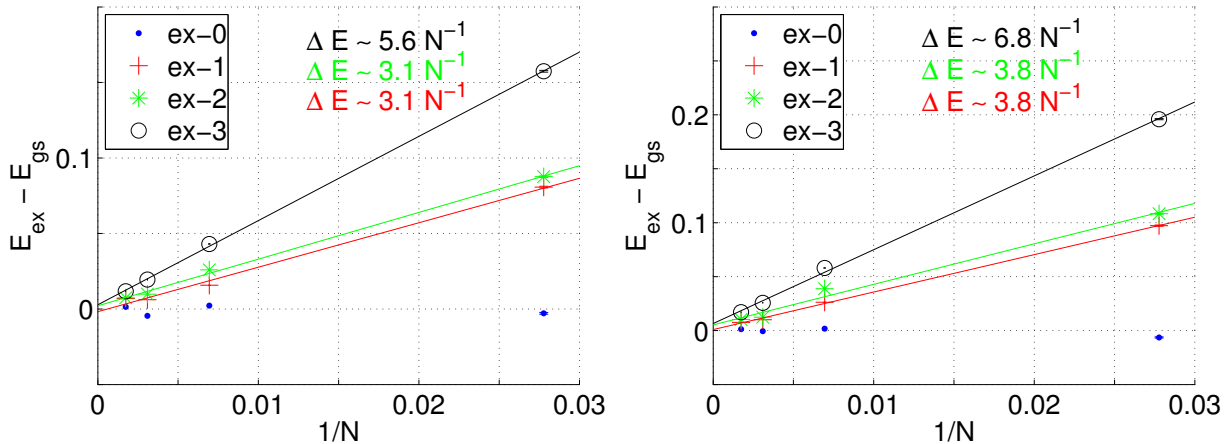

FIG. 9. Excitation energies of (Gutzwiller) projected two-spinon excitations in the Hamiltonians $H^{(1)} = \sum_{\langle i,j \rangle} \boldsymbol{S}_i \cdot \boldsymbol{S}_j$ (left panel) and $H^{(2)} = \sum_{\langle i,j \rangle} P_{ij}$ (right panel). The system sizes are $N = 6 \times 6, 12 \times 12, 18 \times 18$, and $24 \times 24$ sites. See text for the meaning of excitations "ex-0" through "ex-3".

### F. Two-spinon excitations

In Fig. 9, we show the energies of two-spinon excitations of the U(1) Fermi-surface QSL with a large spinon Fermi surface, state (A) in the main text, at 1/3 spinon filling. We consider four types of two-spinon "particle-hole" excitations:

- ex-0: Spinon just below the Fermi surface to just above it.

- ex-1: Spinon just below the Fermi surface to the top of the band.

- ex-2: Spinon from the bottom of the band to just above the Fermi surface.

- ex-3: Spinon from the bottom of the band to the top of the band.

These mean-field excitations are then Gutzwiller-projected to $n_j = 1$ in order to obtain a genuine quantum spin $S = 1$ wave function.

In the left panel of Fig. 9, the energies of these excitations are calculated in a first-neighbor spin $S = 1$ Heisenberg model,

$$H^{(1)} = \sum_{\langle i,j \rangle} \boldsymbol{S}_i \cdot \boldsymbol{S}_j. \qquad (4)$$

In the right panel, the energies are calculated for the model

$$H^{(2)} = \sum_{\langle i,j \rangle} P_{ij}, \qquad (5)$$

where $P_{ij}$ is the exchange operator of sites $i$ and $j$. In terms of spin-1 operators, we have $P_{ij} = \boldsymbol{S}_i \cdot \boldsymbol{S}_j + (\boldsymbol{S}_i \cdot \boldsymbol{S}_j)^2 - 1$. These energies are calculated on $N$-site triangular-lattice clusters for various system sizes $N$. The corresponding variational ground state energies (Gutzwiller projected Fermi sea) are also calculated and subtracted from the excitation energies. In Fig. 9, these energy differences are shown for various system sizes, and we also display linear interpolations in $N^{-1}$.

As the number of sites $N$ goes to infinity, the two-particle excitation energies are expected to collapse to the ground state, and this can indeed be observed from the interpolations in Fig. 9. The excitation energies per site, however, are finite, and they are given by the *slope* of the $1/N$ interpolation. The following can be observed: (i) ex-0 (two-particle excitation close to the Fermi surface) has essentially the same energy as the ground state, (ii) ex-1 and ex-2 have roughly the same energy per site, and (iii) the energy of ex-3 (spanning the full spinon band) is about the sum of the latter two. These observations indicate that a picture of renormalized mean-field energies for two-spinon excitations is appropriate for the U(1) Fermi surface state (A).

As announced in the main text, we extract an excitation energy per site $e_{ex} \simeq 4J$ from these calculations (ex-1 or ex-2). We define this as the experimentally detectable spinon bandwidth. The spectral weight of ex-3 (spinon far below to far above the Fermi surface) is strongly suppressed.

13



\end{document}